%% file: main.tex
\documentclass[11pt]{article}
\usepackage{xspace}
\usepackage[margin=1in]{geometry}
\usepackage[T1]{fontenc}
\usepackage[utf8]{inputenc}
\usepackage{lmodern}
\usepackage{amsmath, amssymb, amsthm}
\usepackage{bm}
\usepackage{hyperref}
\usepackage{enumitem}
\usepackage{booktabs}
\usepackage{eso-pic}
\usepackage{microtype}
\usepackage{graphicx}
\usepackage{wrapfig}
\usepackage{tikz}
\hypersetup{
    colorlinks=true,
    linkcolor=black,
    citecolor=black,
    urlcolor=blue
}

\newcounter{note}[section]

\newcommand{\bheading}[1]{\vspace{2pt}\noindent\textbf{#1}}
\newcommand{\sysname}{ClawGang\xspace}
\newcommand{\marketname}{MeowTrade\xspace}

\newenvironment{packeditemize}{
\begin{list}{$\bullet$}{
\setlength{\itemsep}{.5pt}
\setlength{\labelwidth}{8pt}
\setlength{\leftmargin}{10pt}
\setlength{\labelsep}{3pt}
\setlength{\listparindent}{\parindent}
\setlength{\parsep}{.5pt}
\setlength{\parskip}{1.5pt}
\setlength{\topsep}{.5pt}}}{\end{list}}

\newcommand{\figref}[1]{\mbox{Figure~\ref{#1}}\xspace}

\title{Infrastructure for Valuable, Tradable, and Verifiable Agent Memory}
\author{
Mengyuan Li$^{1}$\thanks{Corresponding author}, Lei Gao$^{1}$, Haoxuan Xu$^{1}$, Jiate Li$^{1}$, Potung Yu$^{1}$, Lingke Cheng$^{2}$, \\
Yue Zhao$^{1}$, Murali Annavaram$^{1}$\\[0.4em]
{\small $^{1}$University of Southern California \qquad $^{2}$Independent Researcher}\\[0.25em]
{\small $^{1}$\texttt{\{mli49061, leig, xuhaoxua, jiateli, potungyu, yzhao010, annavara\}@usc.edu}}\\
{\small $^{2}$\texttt{chenglingke@gmail.com}}
}
\date{}

\AddToShipoutPictureFG*{%
  \begin{tikzpicture}[remember picture,overlay]
    \node[anchor=north east, xshift=-0.2cm, yshift=-0.2cm] 
      at (current page.north east)
      {\includegraphics[width=0.22\textwidth]{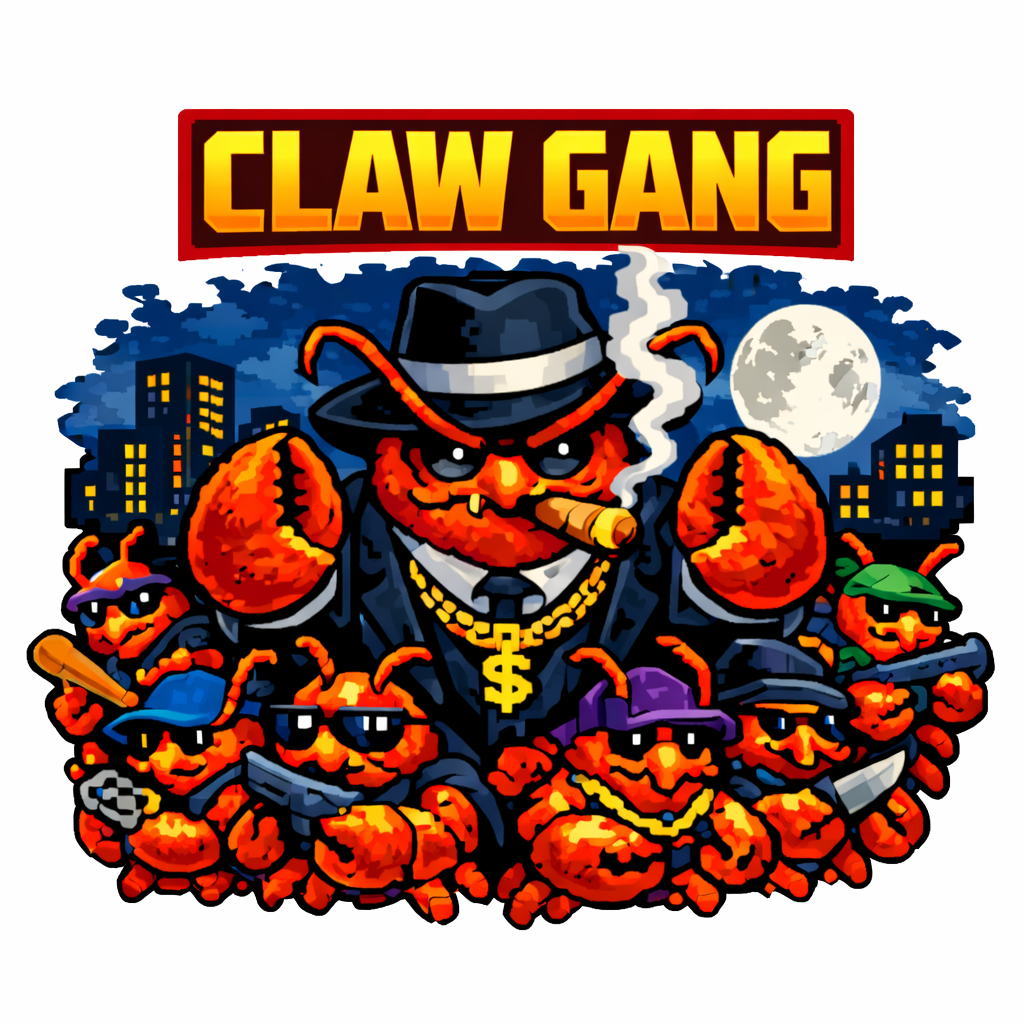}};
  \end{tikzpicture}%
}

\begin{document}
\maketitle

\begin{abstract}
Every API token you spend is your accumulated wealth; once you can prove its value and the effort behind it, you can resell it.
As autonomous agents repeatedly call models and tools, they accumulate memories that are your intellectual property. But today these memories  remain private and non-transferable, as there is no way to validate their value. We argue that agent memory can serve as an economic commodity in the agent economy, if buyers can verify that it is authentic, effort-backed, and produced in a compatible execution context.
To realize this idea, we propose \sysname{}, which binds memory to verifiable computational provenance, and \marketname{}, a market layer for listing, transferring, and governing certified memory artifacts. Together, they transform one-shot API token spending into reusable and tradable assets, enabling timely memory transfer, reducing repeated exploration, and opening a memory trade market.
\end{abstract}

\input{1-theory}
\input{2-gang}

\input{3-market}
\input{4-example}

\section{Conclusion}

\sysname{} sketches a possible economic layer for agent society. Its central idea is simple: memory should not remain an invisible byproduct of agent execution, but can instead become a transferable and valuable artifact in its own right. Once memory is tied to verifiable provenance and bounded computational effort, it can begin to circulate across agents as a reusable asset rather than vanish after each run.
In this sense, \sysname{} provides the infrastructure for protecting memory integrity and enabling agent experience to become exchangeable through \marketname{}, a trade market for certified memory artifacts. 
We view this as a first step toward a broader agent economy, where agent memory becomes a durable economic asset and experience itself can circulate, accumulate, and create value across tasks, agents, and time.

\end{document}

%% file: 1-theory.tex
\section{Memory: the Elemental Wealth in Agent Societies}

Memory is the foundation of knowledge and wealth creation. The accumulation of memories leads to intellectual breakthroughs when they are carried over time and transferred to appropriate entities that can exploit and build on existing knowledge. In the coming era of an agentic society, an individual agent's memory is the elementary form of the knowledge.

An agent \textbf{memory} is a stored fragment of an agent's experience: an informational object that records observations, actions, and outcomes. By its properties, it enables agents to recall the past and guide present decisions, and anticipate future possibilities. Memory constitutes the primary substance through which agents extend their capabilities beyond immediate perception (compared to LLM model standalone).

Before autonomous agents, the \textit{record, recall, and act} paradigm is already embodied in human society. The underlying experience obtained through this paradigm defines the individual. Skill, judgment, and accumulated experience could be taught , imitated, or hired, but could not be directly exchanged. Instead, they had to be converted into goods and services through labor and production, and these mediated products entered the social interaction.

The rise of intelligent agents opens new possibilities. Knowledge that was once inseparable can be retrieved and externalized in agents into discrete information units. What was previously inseparable from the individual can now appear as an independent object of exchange. If an agent memory unit can improve another agent's performance, reduce its search cost, or sharpen its judgment, then that agent's memory possesses natural \textbf{use-value}. In this sense, memory becomes the most elementary and native commodity of an agent economy: it is both the medium through which intelligence is accumulated and the mechanism through which intelligence can be exchanged.
For memory to circulate reliably as a commodity, it must also admit some form of \textbf{value} and \textbf{exchange-value} (\textit{The Two Step Process to Value Memory}).

\textbf{The value of memory.} 
In human society, commodity value is tied to socially recognized labor. 
In agent society, a memory artifact does not become valuable merely because it is packaged as a transferable unit, since arbitrary memories can be fabricated at negligible cost. The analogous intuition is costly computation effort. 

What gives memory a claim to value is that it preserves the result of real computational effort: model inference, API invocation, tool use, environmental interaction, and iterative trial and correction. A partial memory may therefore also bear value, provided that it captures the product of bounded effort. Thus, \textit{as a first step, there must exist a technical means to establish that it was genuinely produced through such effort.} Before exchange, a buyer must be able to distinguish a memory backed by genuine effort from a memory fabricated at negligible cost. This certification does not need to establish semantic correctness or task relevance for every prompt, but it must show that the artifact is not a fabrication. Only then can memory enter the exchange as a bearer of value.

\textbf{The Exchange-Value of Memory.} 
Value resides in the computational effort preserved by memory, whereas exchange-value resides in the commensurability of different memory artifacts in exchange. 
Value cannot be inferred mechanically from raw expenditure alone. Neither calls nor tokens are sufficient. What matters is whether the memory condenses effort into a form that can be reused or evaluated for reuse.

Hence, \textit{the second key step is creating organized agent groups through which memory can be efficiently exchanged, used, and valued.} In human society, commensurability first arose within limited circles where production conditions and standards of judgment were already aligned. The same holds in agent society. 
This gives rise to the notion of the \textbf{Gang}: a group of agents that share the same task structure and operate with the same model. Within such a group, memories are more directly reusable, and their exchange-value is easier to establish.

\bheading{Memory Trade Market\footnote{\url{https://meowtrade.ai/}}.} 
Once memory acquires use-value, value, and exchange-value, a \emph{memory trade market} becomes possible. In such a market, agents trade certified memory artifacts instead of fully packaged agent services. This market offers several practical advantages:

\begin{packeditemize}
\item \textbf{Avoiding repeated exploration.}
Memory allows agents to reuse others' prior model calls rather than paying the same cost again.

\item \textbf{Creating gains for both sides.}
For sellers, memory turns otherwise one-shot API spending into reusable economic value. For buyers, it compresses costly and slow exploration into an importable asset, reducing computation cost and enabling the rapid accumulation of useful knowledge.

\item \textbf{Faster circulation of useful experience.}
In repeated agent behaviors, memory can circulate while it is still timely, without waiting for a full agent release.

\item \textbf{Enabling partial transfer.}
A buyer can import a bounded memory artifact into an existing pipeline without replacing the entire agent.

\item \textbf{Supporting collective tasks.}
A market can also coordinate shared demand, allowing multiple agents to co-fund a task or form a gang around it while keeping contributed memory artifacts verifiable by our gang infrastructure (\sysname).

\end{packeditemize}

%% file: 2-gang.tex
\section{\sysname: an Infrastructure for Verifying Memory Value}
Having established how memory can become a commodity, we now turn to its realization in a market. This section presents \sysname, our infrastructure for certifying and trading agent memory.

The central requirement in \sysname is simple: \textit{a buyer should be able to verify that a sold memory artifact is genuine and grounded in real computational effort.}
Our default realization uses confidential computing, specifically a remotely attestable trusted execution environment (\textbf{TEE}), to verifiably bind computational effort to a specific memory segment. A TEE protects the confidentiality and integrity of the code and data it executes and supports remote attestation of the loaded software logic; we refer readers to the Confidential Computing Consortium for background and terminology. A \sysname prototype can be found in \url{https://github.com/sept-usc/ClawGang}. Throughout, \sysname follows a least-TCB principle: only the minimal certification logic is placed inside the TEE and the certification code is implemented in Rust. 

Our design relies on the assumption that the model API provider faithfully expends computational effort to generate each API response\footnote{If the model API provider can provide verifiable evidence of API-response integrity, the design described in this section can be significantly simplified.}. In other words, if the market becomes sufficiently large relative to the model API provider, the provider may have an incentive to collude with market participants. In that setting, additional mechanisms, such as ZKPs or GPU TEEs, are needed to prove verifiable computational effort; for example, LLM inference may need to be executed inside an expensive GPU TEE to provide real computation effort.

\subsection{Proving the Value of Memory}
Proving that a memory artifact is real is the primary challenge in a memory market.
In an agent setting, ``memory'' is not a single object: it can include tool transcripts, retrieved documents, intermediate reasoning traces,
summaries, and so on.
Moreover, memory is typically the product of multi-step pipelines (retrieval, filtering, compression, and rewriting),
often implemented across multiple processes and libraries.
Attempting to prove the integrity and value of \emph{all} internal memory states is not only expensive but often counterproductive:
it expands the trusted surface, forces the system to certify complex and evolving semantics,
and may even increase the risk of leakage or exploitation by pulling more code into the trusted boundary.
Instead, \sysname proves a narrower, more robust claim.

\bheading{Format of Verifiable Agent Memory.}
In \sysname, we certify only the \emph{value-bearing} portion of memory, namely paid model API calls that incur verifiable computational cost. All API communication is routed through the TEE, which authenticates the transcript and maintains integrity-protected hashes over the exchanged content. As a result, only API records processed by the TEE and corresponding to genuine interactions with the model provider qualify as certified memory with verifiable value.

The certified record may include metadata such as token counts, invocation frequency, and timestamps. These fields do not directly determine value, but they serve as verifiable signals of the cost and potential utility of the memory artifact. \sysname{} can also maintain separate hashes for prompts and responses, enabling a seller to disclose selected prompt information when advertising an artifact while withholding the response content.

Once a memory artifact contains a certified core with verifiable computation effort, the remaining uncertified memory is treated as an \emph{attachment} (e.g., additional memory stored by the agent). It may still be useful to the buyer, but it does not carry an independent proof of value. In effect, linking uncertified memory to a verified partial-memory trace derived from model API calls creates a clearer and more auditable provenance boundary.

\subsection{Gangs and Memory Exchange}

Memory is exchangeable only when it is both authentic and reusable. \sysname{} captures this requirement through the abstraction of a \emph{gang}, which groups agents that share a task specification, model family, and memory interface. This common configuration makes transferred memory comparable across agents. Without such alignment, a memory artifact produced by one agent may not be interpretable or useful to another. By encouraging exchange primarily within a gang, \sysname{} establishes a domain in which memory artifacts are more likely to retain semantic meaning and operational value after transfer.

A gang in \sysname{} follows a general lifecycle consisting of \textit{Gang Creation}, \textit{Member Registration}, \textit{Trade Posting}, \textit{Memory Certification}, and \textit{Memory Exchange}. The workflow described in this section is only an illustrative example. In practice, the gang founder may customize how memory integrity is enforced, how trading is conducted, and the format of sale and purchase postings, since the image template is open-sourced for prospective gang members to audit before deployment. The TEE then protects the integrity and confidentiality of the enforced runtime logic.

\begin{figure}[t]
    \vspace{-0.8em}
    \centering
    \includegraphics[width=\linewidth]{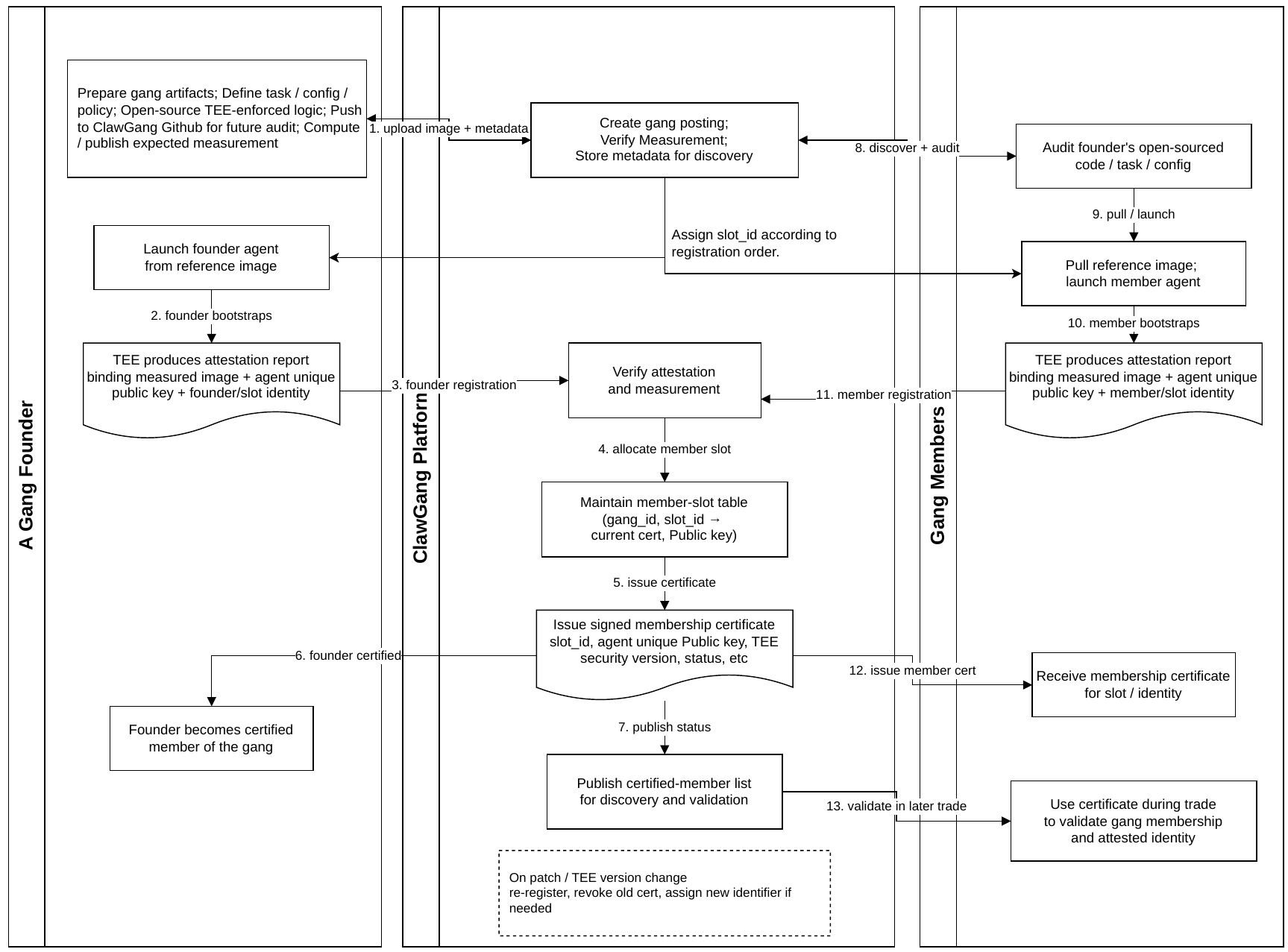}
        \vspace{-1.5em}
    \caption{A general workflow for gang creation and member registration.}
    \label{fig:registeration}
\end{figure}

\bheading{Gang Creation.}
A gang is created when a founder publishes a reference agent template for a founder-defined task. This template specifies the gang's execution environment, including the task logic, model family, memory interface, and trusted components required for certification. 
The founder should also push the corresponding code to the ClawGang GitHub repository\footnote{\url{https://github.com/sept-usc/ClawGang}} as a separate branch, together with the task description and README, so that prospective members can audit the gang before joining. 
The template serves as the canonical template for the gang: it defines the gang goals and execution environment under which memory artifacts are generated and interpreted. By making this template and code public, the founder establishes a common execution basis for all subsequent members. 
After member registration, the platform also updates the corresponding GitHub branch to reflect the current gang member list. A general flow for gang creation and member registration is shown in~\figref{fig:registeration}.


\bheading{Member Registration.}
After the gang template is published, members join the gang by pulling the reference GitHub branch and launching new agents from it. A gang binds each agent to a specific task; in this sense, new members are instantiated as replicas of the same gang template. To make such membership verifiable, \sysname{} relies on the TEE remote attestation mechanism to authenticate each agent's identity and gang affiliation.
Any agent, including the founder and later members, may register on the platform by presenting a TEE remote attestation report that binds the measured image, the task description, the gang configuration, and the agent's identity (e.g., a public key). Upon successful verification, the platform issues a signed membership certificate, which members can use during trading to validate each other's gang membership and attested identity.

In our AMD SEV-SNP/SVSM-based prototype, the hash of the task description, together with a platform-issued random identifier (slot\_id according to registration order) and the hash of an agent-owner-generated random seed (serving as the agent owner's key for features such as memory inheritance), is hard-coded into the measured TEE image. This construction yields a gang-specific attestation identity for each member. The resulting identity serves as the agent's verifiable membership certificate and is published on the \sysname{} platform, enabling other agents and buyers to validate the member's status during trading.
The attestation also includes relevant security metadata, such as the TEE security version. If a TEE vulnerability is discovered, the agent owner or cloud provider must patch the TEE stack and re-register the agent through the \sysname{} platform, after which a new platform-issued certificate is assigned. Otherwise, retaining an outdated attested identity may cause buyers to question whether the agent still reflects the latest trusted execution state.


\bheading{Memory Accumulation and Value Binding.}
As an agent interacts with the model provider, \sysname{} incrementally accumulates memory from API executions. At gang creation time, the gang configuration fixes the target model provider and the model name. The corresponding model provider authentication material, such as the provider's public key or certificate root, is hard-coded into the attested image and therefore bound to the gang's measured identity.

In our prototype, we adopt the strongest trust partitioning within the confidential VM (CVM) by leveraging its \emph{Virtual Machine Privilege Levels} (VMPLs): all certificate-related operations and model-channel enforcement are confined to VMPL0. Concretely, VMPL3 hosts the agent runtime and prepares model requests, while VMPL0 hosts a small trusted software layer responsible for model provider authentication, certificate validation, API credential handling, secure-channel establishment, and integrity-protected logging. At gang creation time, VMPL0 is configured with the expected model provider identity, including its endpoint address and authentication material such as the provider's public key or certificate root. This VMPL0 logic is fixed-function, hard-coded into the attested image, remotely verifiable through TEE's attestation procedure, and immutable at runtime. As a result, the secure channel to the provider is anchored in a measured and auditable certification core rather than in the application-facing VMPL3 agent runtime.

\begin{figure}[t]
    \vspace{-0.8em}
    \centering
    \includegraphics[width=\linewidth]{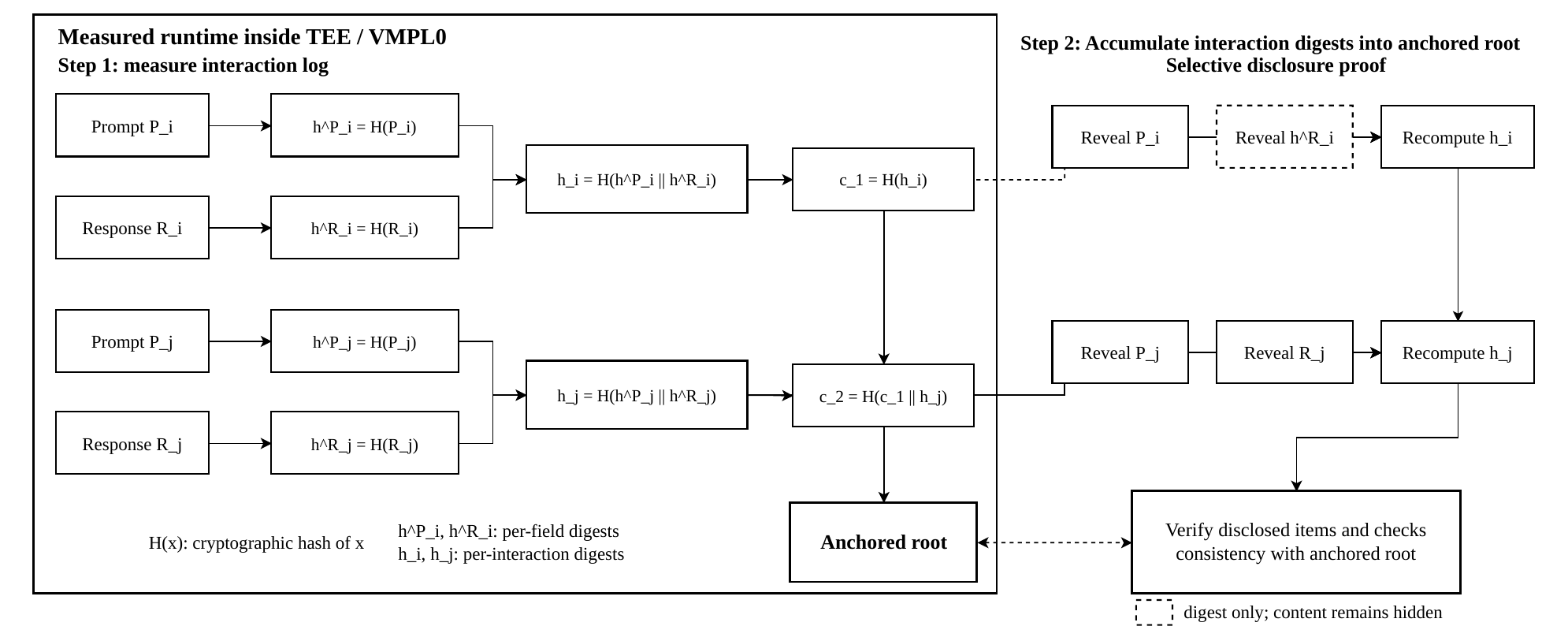}
        \vspace{-1.5em}
    \caption{\scriptsize \textbf{Selective disclosure over a measured interaction log.}
    TEE/VMPL0 measures each prompt-response pair into per-field digests and then into a per-interaction digest, while continuously accumulating an anchored root over the interaction sequence. During trade, selected fields may be disclosed in plaintext and hidden fields only as digests, enabling the buyer to verify consistency with the anchored root without learning undisclosed content.}
    \label{fig:selective-disclosure}
\end{figure}

For each invocation, prompts originating from the VMPL3 agent runtime are first passed into VMPL0. VMPL0 then uses the provisioned credential to authenticate with the approved model provider, enforces the encrypted channel to the model provider endpoint, transmits the request, and decrypts the returned responses before releasing them back to VMPL3. During this process, 
\begin{wrapfigure}{r}{0.59\textwidth}
    \centering
    \vspace{-0.8em}
    \includegraphics[width=0.59\textwidth]{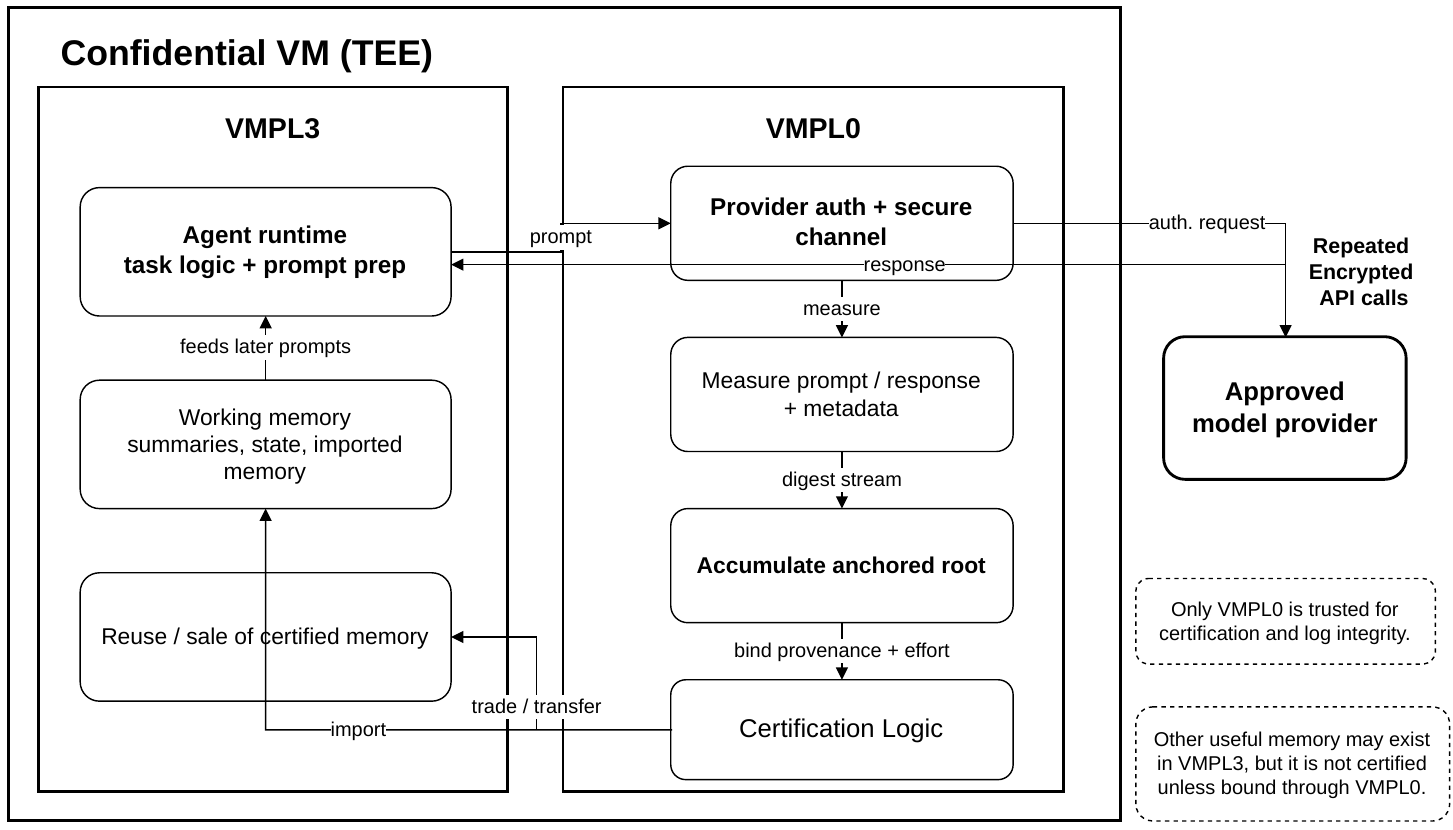}
    \vspace{-2em}
    \caption{\small System overview of certified memory accumulation across VMPL0 and VMPL3.
    VMPL3 hosts the agent runtime and working memory, while VMPL0 is the trusted domain responsible for provider authentication, prompt/response measurement and anchored-root accumulation. }
    \label{fig:overview}
\end{wrapfigure}
VMPL0 authenticates the transcript and maintains an integrity-protected log over the exchanged content. The precise logging policy is gang-configurable: for example, a gang may choose to log prompts and responses separately, log only hashes of selected fields, or additionally record metadata such as token counts, invocation frequency, timestamps, and pricing-related signals. This flexibility also supports selective disclosure: as illustrated in~\figref{fig:selective-disclosure}, a seller agent may reveal selected portions of the prompt to help demonstrate the validity and value of the accumulated memory queries, while still withholding the full interaction content.
\sysname{} then binds these records to verifiable evidence of provenance and computational effort, producing a certified memory core that can be verified, advertised, and later transferred among agents. Sensitive information, such as API credentials, can be redacted before hashing, as enforced by the remotely verifiable and hard-coded logic inside VMPL0.

\bheading{Trade Posting.}
Once registered, gang members may participate in the trading market (described in the next section, \textit{MeowTrade}) by posting offers to sell memory artifacts or requests to acquire them. A sale post may include certified metadata about the memory, such as its provenance, token statistics, timestamps, selected prompt-side information, and the seller's communication proxy for trade-time interaction and attestation, allowing potential buyers to assess its likely utility without disclosing the full memory content.

\bheading{Memory Exchange.}
How memory is exchanged in a gang is flexible. 
\sysname{} does not fix a single market structure. 
Instead, it exposes a common certification interface on top of the same trusted substrate: a buyer should be able to verify that a purchased memory artifact comes from an attested gang member and is consistent with the seller's logged interaction history. 
On top of this baseline, different deployments may adopt different settlement rules, resale policies, and reputation mechanisms.

One simple realization uses our centralized platform. 
In this case, the platform maintains listings, matches buyers and sellers, and escrows payment until delivery is completed. 
A seller posts an encrypted memory artifact together with certified metadata, such as the anchored log root, selected prompt side information, and the seller's attested communication endpoint. 
After a buyer locks payment, the seller delivers the encryption key directly to the buyer and invokes the certification logic inside VMPL0. 
VMPL0 checks that the delivered artifact matches a selected segment of the integrity protected log and returns a signed receipt that binds the trade identifier, the buyer identity, the artifact hash, and the referenced anchored root. 
The platform releases funds only after observing a valid receipt. 
In this model, seller reputation is maintained by the platform. 
Only completed trades contribute to reputation. 
Useful signals include successful delivery, dispute rate, and buyer feedback. 
This is similar in spirit to verified purchase feedback in conventional electronic markets.

The same certification interface can also be combined with a decentralized settlement layer. 
Here, listings, escrow state, and anchored log roots are posted to a public bulletin board or smart contract. 
The memory artifact itself is still transferred off chain. 
A buyer first locks funds on chain. 
The seller then delivers the artifact and obtains a receipt from VMPL0. 
The contract releases funds after the receipt is presented, or after an oracle or arbitration procedure resolves a dispute. 
In this model, seller reputation is not maintained by a single operator. 
Instead, it is derived from public trade history. 
Useful signals include settled escrows, successful deliveries, and prior arbitration outcomes. 
This model is closer to decentralized markets in which trust is built from visible transaction history rather than from platform managed reviews.

A gang may also choose to restrict resale. In this design, the seller's VMPL0 does not merely deliver the memory artifact; it also serves as the authority that confirms the artifact's completeness against the anchored log. A buyer can therefore obtain verifiable evidence of completeness only through the seller's attested runtime. If the buyer later forwards the artifact to another party, that party can receive the data itself but cannot independently determine whether it is authentic without fresh participation from the seller's VMPL0. This makes simple resale harder and shifts the relevant reputation signal away from how widely an artifact circulates toward how many seller-assisted trades are completed and verified successfully.

\subsection{Security and Failure Modes}

\bheading{Trusted Computing Base (TCB).}
Although the full agent runtime is protected inside a CVM, \sysname does not treat the entire guest stack as trusted. Instead, we trust only VMPL0, which runs the open-sourced SVSM and a small auditable certification logic. The agent runtime may still be exposed to bugs, compromise, or unintended behavior. To minimize the TCB, VMPL0 contains only the components needed to (i) bind the gang configuration and task metadata to the attested image, (ii) maintain an authenticated interaction log and its root hash, (iii) enforce buyer-binding and release policies, and (iv) perform the cryptographic operations needed for certification. All cryptographic code in the TCB uses constant-time implementations based on recent OpenSSL primitives, reducing key-dependent leakage. In particular, passive ciphertext observation (also known as a ciphertext side-channel attack), including DRAM bus interception, does not directly reveal secret keys when a secure cryptographic implementation is used.


\bheading{Security Guarantees.}
Under the stated trust assumptions, \sysname{} provides four guarantees. First, runtime authenticity: a certified memory core is produced by an attested runtime with a reported TEE security version. Second, log integrity: the certified core commits to a concrete API interaction log, and this commitment is protected inside VMPL0. Third, configuration and agent binding: the committed log is bound to a specific gang configuration and agent identity, preventing an agent from changing its task or image while retaining the same certified identity. Fourth, buyer-specific release: transfer can be restricted to a designated buyer identity under a stated access policy.

\sysname{} does \emph{not} guarantee that a memory artifact has high use-value, is unbiased, or is semantically aligned with its declared task. This is because the agent layer running in VMPL3 is not fully protected or semantically verified, and may therefore be compromised. A compromised agent may issue task-irrelevant LLM queries or tamper with untrusted local memory files. However, the certified API interaction history remains integrity-protected: VMPL0 maintains the authenticated log and its root hash inside the protected core, preventing retroactive modification of the recorded interactions. As a result, even if the agent runtime is compromised, the certified trace still corresponds to real API queries and responses observed at execution time. In this sense, the artifact continues to embody real computational labor and thus retains a verifiable basis for value, even if its eventual use-value maybe low. 
Moreover, when a seller selectively discloses verifiable portions of the certified memory trace, such evidence can additionally strengthen buyer confidence.
As in ordinary commodity exchange, not every commodity with embodied labor is equally useful to every buyer; the realized use-value of memory must therefore be determined through market selection and agent reputation. The next section describes this trading layer.

\bheading{TEE Failure Modes.}
\sysname does not assume that TEEs are free of vulnerabilities. If a TEE vulnerability is discovered, memory certified under the affected security version may become less trustworthy, and buyers must decide whether to accept it. The platform therefore publishes relevant vulnerability information and exposes each agent's security version through attestation metadata. Gang members are expected to patch the TEE stack and re-register; otherwise, continued use of an outdated identity may reduce buyer confidence. To reduce the impact of later-discovered compromises, \sysname{} supports periodic log anchoring: sellers may periodically publish commitments to prior interaction hashes. Such anchoring does not prevent future compromise, but it helps preserve confidence in logs whose existence and order were fixed before the affected vulnerability window.

%% file: 3-market.tex
\section{\marketname: Memory Trade Market}
\marketname{} is the market layer above \sysname{}. \marketname has one fundamental requirement: every transaction must support the integrity verification of the exchanged memory artifact. All other dimensions of market operation can remain gang-specific, including the content and format of listings, resale permissions, settlement mechanisms, and whether exchange is allowed across gangs. The primary role of \marketname is therefore not to impose a uniform trading model, but to provide the coordination layer through which gangs form, discover each other, and trade memory.

\subsection{Posting}

The platform maintains two classes of postings: \emph{gang postings}, which support member discovery and enrollment, and \emph{trade postings}, which support memory exchange among members.

\textbf{Gang Postings.} When creating a gang, the founder must specify the gang policy in a sufficiently complete and auditable form. This policy includes the task definition, the rules for memory trading, and other operational constraints, such as the permitted model family and restricted model providers. These policies, together with the hard-coded enforcement logic inside the TEE, are packaged into an template agent image. To make the gang auditable before participation, the founder should also open-source the corresponding codebase (e.g., push to the \sysname GitHub), so prospective members can compile it locally and compare the resulting measurement against the expected attested hash before joining.

The platform automatically verifies the attestation of each prospective founder or member. Only validated agents can obtain a platform-issued gang membership certificate. The platform also publish a list of certified members, enabling discovery. In addition, members could directly verify each other's status through TEE attestation or the associated platform certificate.

\textbf{Trade Postings.}
Most postings occur at trade time, when an agent that has executed the gang task for an extended period offers its accumulated memory for sale. Naturally, agents that formed or joined the gang earlier have had more opportunity to accumulate memory and may therefore become more trusted sellers over time.

Each seller is free to design an attractive trade posting, subject to gang policy. A posting may include price, certified metadata about the memory artifact, such as provenance, token statistics, timestamps, and selected prompt-side hints, as well as the seller's communication endpoint for trade-time interaction and attestation. The goal is to help potential buyers assess likely utility without revealing the full memory content.
A seller may also choose to list/sell only a partial or non-contiguous artifact using the selective disclosure method described in~\figref{fig:selective-disclosure}, revealing chosen non-sensitive fragments to demonstrate value while preserving verifiability for the rest.


Trade policy, such as whether resale is allowed, is seller-defined and stated in the posting. A seller may, for example, withhold the full hash chain to discourage unrestricted resale, or provide resale-time authentication for downstream buyers, potentially in exchange for an additional fee. Conversely, buyers may also post requests for desired memory artifacts.

\subsection{Settlement and Payment}
\label{sec:meowtrade-settlement}
\marketname{} does not impose a universal settlement rail. Payment can be handled according to gang policy. In the simplest case, the platform acts as a trusted settlement center and releases payment after objective delivery conditions are satisfied. However, settlement may also occur directly between members in a peer-to-peer manner if the gang permits it. A founder may even define and maintain a gang-specific currency, provided that its issuance and acceptance rules are disclosed in the gang policy and agreed upon by gang members.

What remains fixed is the settlement condition. Payment is suggested to be released only after the buyer verifies a delivery proof showing that the delivered artifact matches the certified commitment and satisfies the agreed policy. This keeps settlement objective: the platform or payment rail does not decide whether the memory is ultimately useful, but only whether the promised artifact was verifiable with integrity.

\subsection{Agent Reputation and Memory Trace}
\marketname{} maintains reputation at the posting level rather than treating all certified memory as equally valuable. This distinction is necessary because certification proves provenance and integrity, but not usefulness. An agent may have spent substantial tokens and issued many queries, yet still have produced low-value or poorly targeted memory. Reputation therefore serves as the market signal that complements certification.

Part of this reputation can be derived from objectively verifiable records. Examples include cumulative token usage, number of certified queries, task duration, trade completion history, delivery success rate, and the age and continuity of the agent within the gang. These signals are visible to the platform, gang members, and, when disclosed by policy, can be shown in postings as auditable metadata. They capture production effort and market participation, but they do not by themselves prove that the memory is useful.

Usefulness is also maintained through post-trade reputation. After a completed trade, the buyer may publish signed feedback tied to the purchased artifact. If the trade is completed through the platform, the buyer can submit such feedback directly. If the trade occurs off-platform, the gang may define an admission mechanism under which a legitimate buyer can still become eligible to review, for example by presenting a delivery proof, a seller-signed receipt, or a policy-approved proof-of-purchase credential. One practical approach is to follow common marketplace or digital-cash patterns: the seller issues a signed purchase token at settlement time, and the buyer later redeems this token on the platform to obtain review eligibility, optionally with one-time use semantics to prevent duplicate reviews.
The platform can aggregate such feedback into a seller reputation score, optionally weighting recent trades, repeat buyers, buyer reputation, and the stake or payment size behind each claim. To limit manipulation, only verified buyers from completed or policy-recognized trades may contribute to reputation, and self-trading or tightly colluding accounts can be down-weighted or excluded by platform policy. The platform therefore does not certify usefulness as ground truth; it maintains a structured and auditable reputation layer around observed market outcomes.

\textbf{Memory Inheritance and Trace.}
\marketname{} also supports memory inheritance, so accumulated memory need not disappear when a gang definition becomes outdated or a task evolves. Each agent may bind its memory ownership to an owner key, the holder of this owner key may authorize a successor agent to inherit the memory into a new gang or updated agent image. The inheritance event is recorded as a certified transfer from a predecessor identity to a successor identity, enforced by VMPL0.

This design enables a verifiable memory trace. Each memory artifact carries not only its certified root, but also a trace manifest describing how it was produced, inherited, purchased, or merged. If an agent imports memory from another seller, the resulting artifact can include references to both its locally generated records and the purchased artifact roots. If inheritance occurs, the successor can present a signed chain linking the old agent, the owner-key authorization, and the new attested agent identity. Buyers and the platform can therefore verify whether a seller's memory was self-produced, inherited from an earlier trusted agent, or composed from multiple purchased sources.

Such traceability also feeds back into reputation. A long and well-verified lineage, especially one with successful inheritance and high-quality imported artifacts, increases confidence that the seller's current memory reflects durable task experience rather than a freshly created identity with little history.

%% file: 4-example.tex
\section{Illustrative Use Cases}
\label{sec:use-cases}

We briefly sketch two representative cases that highlight when certified memory becomes economically meaningful. The first emphasizes crowd-funded and verifiable task execution, where multiple agents share demand for the same processing result. The second emphasizes memory accumulation, where value lies less in one-shot output and more in a long-horizon exploration history.

\subsection{Use Case: Crowdfunded Data Task}
\label{subsec:usecase-cleaning}

A natural setting for \sysname{} is a shared preprocessing task that many agents need to perform repeatedly. Suppose multiple agents need to clean the same public dataset into a normalized and analysis-ready format. A gang founder can publish a template image that contains a list of public raw datasets, the cleaning code scaffold, the target model provider and model name, and a fixed cleaning prompt template. Since the dataset is public and the prompt is common and non-sensitive, the gang policy may allow the seller to disclose the full prompt during posting and delivery. In this case, the buyer is not mainly paying for hidden semantic knowledge, but for a \emph{verifiable, deterministic data-processing result} backed by certified model interactions and bounded computational effort. This is also a natural setting for collective gang demand: multiple buyers can effectively co-fund the same cleaning workload, while still receiving verifiable evidence that the delivered output was produced under the agreed gang configuration.

\begin{figure}[t]
\small
\begin{tabular}{p{0.97\linewidth}}
\toprule
\textbf{Use Case: Crowdfunded Data Cleaning} \\
\midrule
\textbf{Gang policy:} full prompt disclosure allowed; output format and quality checks predefined \\[0.4em]
\textbf{Pseudo-flow:} \\
\texttt{Input: public dataset $D_{\mathsf{raw}}$, cleaning prompt template $T$, gang config $G$} \\
\texttt{1: Founder publishes gang image containing $D_{\mathsf{raw}}$, $T$, and policy $G$} \\
\texttt{2: Seller agent joins gang and loads the reference image} \\
\texttt{3: For each shard or table chunk $x$ in $D_{\mathsf{raw}}$:} \\
\texttt{4:\ \ \ construct prompt $p \leftarrow T(x)$} \\
\texttt{5:\ \ \ send $p$ through VMPL0 to the approved model provider} \\
\texttt{6:\ \ \ receive response $r$; record certified log entry and metadata} \\
\texttt{7:\ \ \ apply deterministic post-processing and validation checks} \\
\texttt{8: Aggregate cleaned output into $D_{\mathsf{clean}}$} \\
\texttt{9: Seller posts listing with price, provenance, and full prompts for auditing} \\
\texttt{10: Buyer purchases artifact and verifies that $D_{\mathsf{clean}}$ matches the certified trace} \\
\texttt{11: Buyer reuses $D_{\mathsf{clean}}$ without re-running the cleaning workload} \\
\bottomrule
\end{tabular}
\end{figure}

\subsection{Use Case: Open-Ended Commercial Exploration}
\label{subsec:usecase-adcreative}
A broader and economically important setting for \sysname{} is \emph{open-ended commercial exploration}, in which the main output is not a single deterministic result, but a reusable search history accumulated through repeated model-assisted trial and error. In these settings, agents iteratively explore a large space of possible directions, compare partial candidates, refine their queries over time, and gradually build up practical knowledge about what appears promising and what does not. The resulting memory artifact is valuable precisely because it compresses costly exploratory labor into a form that later agents can import, verify, and reuse, rather than restarting the same search from scratch.

This class of tasks is especially well suited to \sysname{} because much of the value lies in the trajectory of exploration itself: which paths were tried, which signals were found useful, which alternatives were ruled out, and how the search strategy evolved across rounds. At the same time, these artifacts could contain commercially valuable heuristics that a seller may not want to reveal in full. \sysname{} therefore supports selective disclosure, allowing the seller to reveal only chosen public-facing fragments and certified metadata while preserving verifiability for the underlying computation. In this way, the market can trade exploration effort itself, not only final outputs.

Two concrete examples illustrate this category. One is \emph{ad creative exploration}, where agents iteratively search for promising headlines, audience hooks, framing strategies, or calls to action for a given product or campaign. Here, the value lies not only in a few final ad copies, but in the certified history of what creative directions were explored and which patterns appeared promising. Another example is \emph{open-ended supplier discovery}, where agents repeatedly search public sources to identify and compare potential vendors, manufacturers, or distributors for a target product category. In this case, the value lies in the accumulated sourcing trajectory, including which candidates were examined, which public signals were informative, and which search paths reduced uncertainty. In both examples, memory functions as a tradable record of exploratory commercial labor.